\documentclass[twocolumn]{aastex631}
\usepackage{enumitem}
\usepackage{graphicx,url}
\usepackage{color}
\usepackage{hyperref}
\usepackage{amsmath}
\usepackage{amssymb}
\usepackage{mathtools}
\usepackage[htt]{hyphenat}

\defcitealias{Zdziarski2021a}{Z21}
\defcitealias{Garcia2018a}{G18}

\def\j1550{XTE~J1550$-$564}

\def\u1630{4U~1630$-$47}
\def\j1752{XTE~J1752$-$223}
\def\u1543{4U~1543$-$47}
\def\v4641{V4641~Sgr}

\def\swift{{\it Swift}}
\def\nustar{{\it NuSTAR}}
\def\nicer{{\it NICER}}
\def\chandra{{\it Chandra}}

\def\integral{{\it INTEGRAL}}

\def\bhlmxb{BH-LMXBs}

\accepted{by ApJ on Feb 27 2025}

\shorttitle{\v4641 in its soft spectral state}
\shortauthors{Connors et al.}

\setlist[itemize]{leftmargin=*}

\begin{document}

\title{Reprocessing from highly ionized gas in the soft spectral state of \v4641 with \textbf{\textit{NuSTAR}}}


\correspondingauthor{Riley~M.~T.~Connors}
\email{riley.connors@villanova.edu}

\author{Riley~M.~T.~Connors}
\affiliation{Department of Physics, Villanova University, 800 Lancaster Avenue, Villanova, PA 19085, USA}
  
 \author{Joey~Neilsen}
\affiliation{Department of Physics, Villanova University, 800 Lancaster Avenue, Villanova, PA 19085, USA}

\author{Aarran~W.~Shaw}
\affiliation{Department of Physics and Astronomy, Butler University, 4600 Sunset Ave, Indianapolis, IN 46208, USA}

\author{James~F.~Steiner}
\affiliation{Harvard-Smithsonian Center for Astrophysics, 60 Garden St., Cambridge, MA 02138, USA}

\author{Federico~Vincentelli}
\affiliation{Instituto de Astrofisica de Canarias, E-38205, La Laguna, Tenerife, Spain}
\affiliation{Departamento de Astrofisica, Universidad de La Laguna, E-38206 La Laguna, Tenerife, Spain}

\author{Javier~A.~Garc\'ia}
\affiliation{NASA Goddard Space Flight Center, 880 Greenbelt Rd, Greenbelt, MD 20771, USA}
\affiliation{Cahill Center for Astronomy and Astrophysics, California Institute of Technology, 1216 E California Blvd, Pasadena, CA 91125, USA}

\author{Phil~Uttley}
\affiliation{Anton Pannekoek Institute, University of Amsterdam, Science Park 904, Amsterdam, Netherlands}

\author{Ron~Remillard}
\affiliation{MIT Kavli Institute for Astrophysics and Space Research, 70 Vassar St., Cambridge, MA 02139, USA}

\author{Guglielmo~Mastroserio}
\affiliation{Dipartimento di Fisica, Universit\'a degli Studi di Milano, Via Celoria 16, I-20133 Milano, Italy}
  
\begin{abstract}
\v4641\ is a low-mass black hole X-ray binary system with somewhat puzzling spectral characteristics during its soft state. Recent high-resolution spectroscopic studies of \v4641\ have revealed strong ionized emission line features in both the optical and X-ray bands, including P-Cygni signatures, and an unusually low soft state luminosity, indicating that the central engine is obscured. Here we present an analysis of five \nustar\ observations of \v4641\ taken during its recent outburst in 2021, when the source was in the soft state. We identify highly ionized Fe K emission lines, consistent with a combination of the near-neutral $6.4$~keV Fe K$\alpha$ line, and the H-like and He-like Fe K$\alpha$ and Fe K$\beta$ transitions found at $6.7\mbox{--}7$~keV and $\sim8$~keV, and find no evidence for strong relativistic broadening. The line fluxes correlate linearly with the observed disk continuum flux, implying a direct connection between the central engine and the reprocessing region. Most interestingly, all five spectra also show a persistent highly ionized Fe K continuum edge feature at $\sim9$~keV with a stable optical depth, which is likely smeared, implying a localized reprocessing zone. We find tentative supporting evidence for obscuration of the inner accretion disk based on its unusually low intrinsic luminosity, however, the \nustar\ spectra do not require obscuration from cold, optically thick gas. 

\end{abstract}


%
%
%
\section{Introduction}\label{sec:intro}

Galactic black hole (BH) transients are X-ray binary systems formed of a BH and a stellar companion. BH binaries in which the stellar companion is low mass are referred to as BH low-mass X-ray binaries (BH-LMXBs). These are evolved binary systems with orbits tight enough such that the companion star is filling its Roche lobe, causing matter to transfer onto the BH. This process, called accretion, results in the release of powerful emission, since a significant fraction of the available gravitational energy is converted into thermal and kinetic energy in the accreting gas. BH-LMXBs spend the majority of their lifetimes in periods (years to decades) of quiescence, dim states in which the accretion rate is low, and thus the radiative luminosity (or power) is many orders of magnitude below the Eddington limit---the theoretical upper bound on luminosity imposed by the competing gravitational and radiative forces. When outbursts occur, BH-LMXBs typically increase in luminosity by several orders of magnitude on timescales of days to weeks. They often present with distinct spectral and time-variability properties in the X-ray band which has been the basis for classifying distinct spectral states, broadly termed `hard' and `soft', with intermediate sub-divisions (see, e.g., \citealt{Homan2005,Remillard2006}). 

During hard spectral states the broadband X-ray spectra of \bhlmxb\ are power-law distributed, with photon index $\Gamma\sim1.5\mbox{--}2$, and peak X-ray luminosities of $\sim10\%~L_{\rm Edd}$ \citep{Remillard2006,Done2007}. When transitioning to the soft spectral state, where a blackbody disk continuum instead dominates the spectrum, \bhlmxb\ can reach luminosities $L_{\rm X}\ge L_{\rm Edd}$. Soft states are also notable for the weakness or absence of the strong fast variability and compact radio jet emission that characterize hard states in \bhlmxb\ (see, e.g., \citealt{MunozDarias2011,Fender2014}). Instead, the soft state is often accompanied by strong outflows that are termed accretion disk winds, distinct from jets in that they likely originate from the outer regions of the accretion flow, and are slower and less collimated than jets (see, e.g., \citealt{Ponti2012}). 

Signatures of disk winds are typically detected spectroscopically, from near-infrared through to X-ray wavelengths, as ionized absorbers. These winds are seen in high inclination systems only, and thus likely have an equatorial geometry, such that we can only detect their absorption signatures when the accretion flow is edge on \citep{DiazTrigo2006,Miller2006,Ponti2012}. Earlier wind studies indicated that disk winds are preferentially observed during the soft state, and thus that winds appear when the radio jet is suppressed \citep{Miller2006,Neilsen2009,Ponti2012,Ponti2016}, however, later studies have shown that winds can still appear alongside the jet \citep{King2015,MunozDarias2016,Homan2016}. 

The primary X-ray spectral identifiers of disk winds are blue-shifted absorption lines (via highly ionized H-like and He-like transitions; \citealt{Lee2002,Ueda2004,Miller2006,Miller2008}). High-resolution spectroscopic observations of these winds, in both the optical and X-ray wavebands, allow us to measure various wind properties, in particular to derive the outflow velocity. It is also possible in some cases to identify emission that is scattered out of our line of sight via the detection of strong emission lines with clear P-Cygni profiles \citep{Brandt2000,King2015,MunozDarias2016,MunozDarias2018}. Such measurements suggest a spherical geometry for the obscuring material, such that we are able to observe reprocessed emission from the obscured central source of photons, i.e., the inner accretion flow. \v4641, the BH-LMXB that is the focus of this study, is one such source in which emission line features have been observed and analyzed in detail via high-resolution spectroscopy in both the optical and X-ray bands \citep{MunozDarias2018,Shaw2022}. 

\subsection{\v4641}
Optical studies in quiescence show the central compact object of \v4641\ is a BH of mass $M_{\rm BH}=6.4\pm0.6~M_{\odot}$, with a B9III (post main-sequence) companion star of mass $M_2=2.9\pm0.4~M_{\odot}$, at a distance of $D=6.2\pm0.7$~kpc, binary inclination $i=72\pm4$~deg, and with an orbital period of $\sim2.8$~days \citep{Orosz2001,MacDonald2014}. \v4641\ is in a unique position as a BH-LMXB due to several of its behavioral characteristics. Since its discovery as a bright Galactic X-ray source in 1999 \citep{in'tZand1999,in'tZand2000}, it has undergone outbursts that vary significantly in duration and peak luminosity with a recurrence time of $\sim220$~days \citep{Tetarenko2016}. 

The 1999 outburst of \v4641\ was both rapid and bright, with the X-ray flux rising to $>12$~Crab within just $8$~hours, before fading over two orders of magnitude down to $\sim0.2$~Crab within $2$~hours \citep{Smith1999,Hjellming2000}. Further analysis of the available data suggested that the source was in a super-Eddington state during this time, and that the disk continuum emission may be obscured, enshrouded in an envelope \citep{Revnivtsev2002}. Subsequent outbursts have been comparatively stunted, with longer, fainter periods of X-ray activity. Furthermore, \v4641\ is seen to remain in a soft, disk-dominated spectral state down to much lower Eddington-scaled luminosities (as low as $6\times10^{-4}~L_{\rm Edd}$; \citealt{Bahramian2015}) than typical \citep{Dunn2010}. 

Analysis of \chandra\ High Energy Transmission Grating (HETG) spectra from observations of the 2020 outburst of \v4641\ revealed a rich set of ionized emission lines atop a strongly scattered disk continuum, and showed that there is likely more than one photoionization region surrounding the central engine \citep{Shaw2022}. Further analysis, together with spectral data from the {\it Nuclear Spectroscopic Telescope Array} (\nustar; \citealt{Harrison:2013}), demonstrated that the central engine is obscured. The low energy coverage of \chandra\ required a partial covering gas to be adequately modeled, with some evidence for weak P-Cygni line profiles (blue absorption and red emission components). These factors led \cite{Shaw2022} to conclude there may be a spherical geometry to the surrounding material, though they did not specifically model this geometry, and the blue-shifted absorption features were not detected with high statistical significance.

\v4641\ underwent an X-ray brightening in 2021, and was detected by the {\it Monitor of All-Sky X-ray Image} ({\it MAXI}; \citealt{Matsuoka2009}) Gas Slit Camera (GSC; \citealt{Mihara2011}) alert system in October 2021 \citep{Negoro2021ATel_v4641}. It was shown to have begun rising in flux as early as August 2021, and had reached a $2\mbox{--}10$~keV flux of $\sim20$~mCrab as of October 12 2021. Follow up observations with the {\it INTErnational Gamma-Ray Astrophysics Laboratory} (\integral; \citealt{Winkler2003}), the {\it Neil Gehrels Swift Observatory} (\swift; \citealt{Gehrels2004}) X-ray Telescope (XRT; \citealt{Burrows2005}) and the {\it Neutron Star Interior Composition Explorer} (\nicer; \citealt{Gendreau2012}), confirmed the X-ray brightening \citep{Motta2021ATel_v4641}. Further \nicer\ observations on October 13, 14 and 16 revealed more details about the source \citep{Buisson2021ATel_v4641}: \v4641\ was varying slightly in flux day-to-day, between $\sim15$ and $\sim20$~mCrab, and showed a strong spectral turnover in the $2\mbox{--}10$~keV band of \nicer, consistent with a soft thermal disk blackbody spectrum with inner disk temperature $kT_{\rm in}\sim1.3$~keV. Most interestingly, \nicer\ detected prominent emission line features in the iron K band of the X-ray spectrum ($6.7\mbox{--}7$~keV). \cite{Buisson2021ATel_v4641} reported that by stacking the \nicer\ spectra they were able to detect a double-peaked line shape in the iron K band, with centroids at $\sim6.7$~keV and $\sim6.97$~keV, which corresponds to the He-like and H-like ionized Fe K$\alpha$ lines (\ion{Fe}{25} and \ion{Fe}{26}). Furthermore, they report that they do not detect statistically significant changes to the line properties (across the three \nicer\ observations, a 4-day window), despite moderate variations in the X-ray continuum, potentially indicating some evolution in either the geometry or ionization structure of the emission line region. 

\begin{figure}
\includegraphics[width=\linewidth]{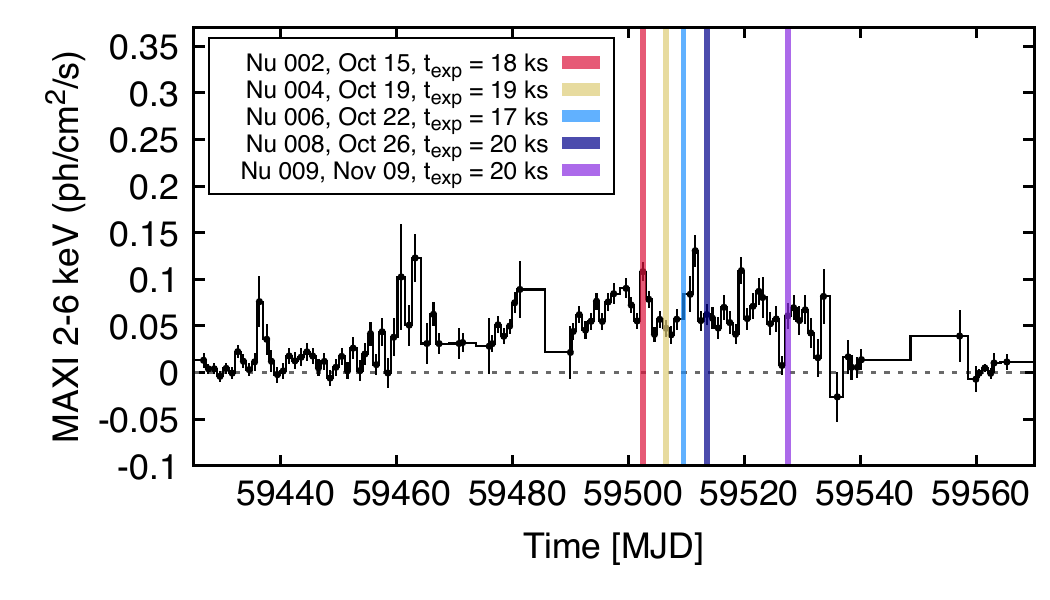}
\includegraphics[width=\linewidth,trim={0 50 0 0}]{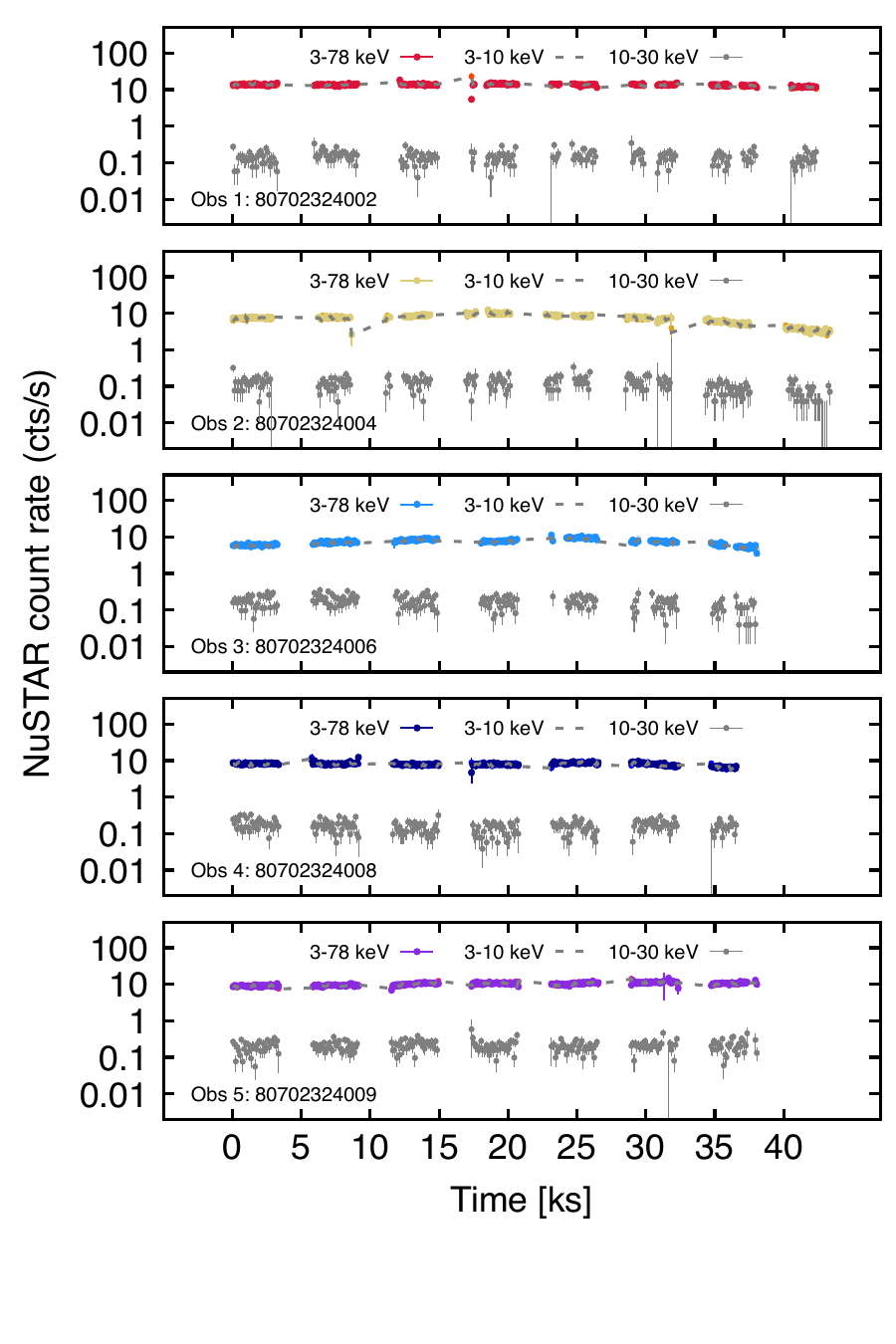}
\caption{{\bf Top}: the long term MAXI light curve of \v4641. The colored bars indicate the \nustar\ observation times associated with our study, with ObsID 80702324. {\bf Bottom}: the five individual \nustar\ light curves, in $100$s bins. Each panel shows the color-coded $3\mbox{--}78$~keV rates, plus the soft and hard band count rates in gray dashed lines and points respectively, showing that the source is dominated by variable (long-term) soft emission. }
\label{fig:maxi_lc}
\end{figure}

\nustar\ initially observed \v4641\ on October 15 2021 as part of a Target-of-Opportunity (ToO) Guest Observer (GO) program. The source was observed a total of five times over a month-long period (see Figure~\ref{fig:maxi_lc}): the first four observations were spaced a few days apart and spanned $\sim2$~weeks before a $\sim2$~week gap, upon which the fifth snapshot was taken. Each snapshot lasted $\sim20$~ks. 

\subsection{This Work}
In this paper we present broadband X-ray spectral modeling of the five \nustar\ observations of \v4641 in October-November 2021, during which the source was in a soft spectral state, with a bright disk component and a faint power law tail, as well as several ionized emission lines and an apparent absorption edge feature at $\sim9\mbox{--}10$~keV. In Section~\ref{sec:data} we discuss the data reduction and analysis, in Section~\ref{sec:modeling} we present the results of comprehensive spectral modeling of all five \nustar\ observations of \v4641, and in Section~\ref{sec:discussion} we discuss the implications of our results, with a particular focus on the ionized Fe K complex and the disk continuum constraints. 

\section{Observations and Data Reduction}\label{sec:data}

The long-term MAXI light curve of \v4641\ is shown in Figure~\ref{fig:maxi_lc}, with the five \nustar\ observation periods overplotted, and individual \nustar\ light curves in sequential panels. The ObsID is 80702324 with appended Sequence IDs 002, 004, 006, 008, and 009. We have included full band ($3\mbox{--}78$~keV), soft band ($3\mbox{--}10$~keV) and hard band ($10\mbox{--}30$~keV) light curves, which demonstrate that the source is in a soft state with power law flux on the few \% level. A rudimentary check of the short term variability of the source (by generating power spectra from all 5 observation light curves) reveals only Poisson noise, i.e., there is no short term variability across the entire \nustar\ passband.

We performed data reduction on all five \nustar\ observations in an identical manner. All data processing was performed using \texttt{HEASoft v6.30.1}, \texttt{nustardas v2.1.2} and the most recent \nustar\ calibration database (CALDB) at the time of analysis, v20221019 released on October 19 2022. Initially we processed all five observations using \texttt{nupipeline} with default settings. Despite \v4641\ being a bright Galactic X-ray source, we did not process the data using the bright status flag\footnote{ \url {https://heasarc.gsfc.nasa.gov/docs/nustar/nustar_faq.html} } since a rudimentary scan of the light curves (in 1s bins) showed the total observed count rate consistently remained below $100~{\rm cts~s^{-1}}$.

We then used \texttt{DS9 SAOIMAGE} \citep{ds9} to select the source and background regions prior to extracting data products. \nustar\ has two focal plane modules (FPM A and B) each comprised of four $32\times32$ pixel detectors arranged in a $2\times2$ square grid (Det 0, Det 1, Det 2, and Det 3, clockwise). A prominent stray light source is present in both FPMs, and sits on the same detector (Det \#0) upon which the source is located.~We therefore extracted the background spectrum from Det \#2. We extracted source spectra and light curves for FPMs~A and B (using \texttt{nuproducts}) from a circular region of 60" centered on the source, and selected a background of the same size in our chosen off-source region. We group all data with a minimum $S/N=5$, and only include data between $3$ and $79$~keV. In the analysis that follows we adopt Cash statistics throughout, given the low count rates at high energies (see Figure~\ref{fig:spectrum}). 

\section{Spectral Modeling} \label{sec:modeling}

\begin{figure}
\includegraphics[width=\linewidth,trim={0 15 0 0}]{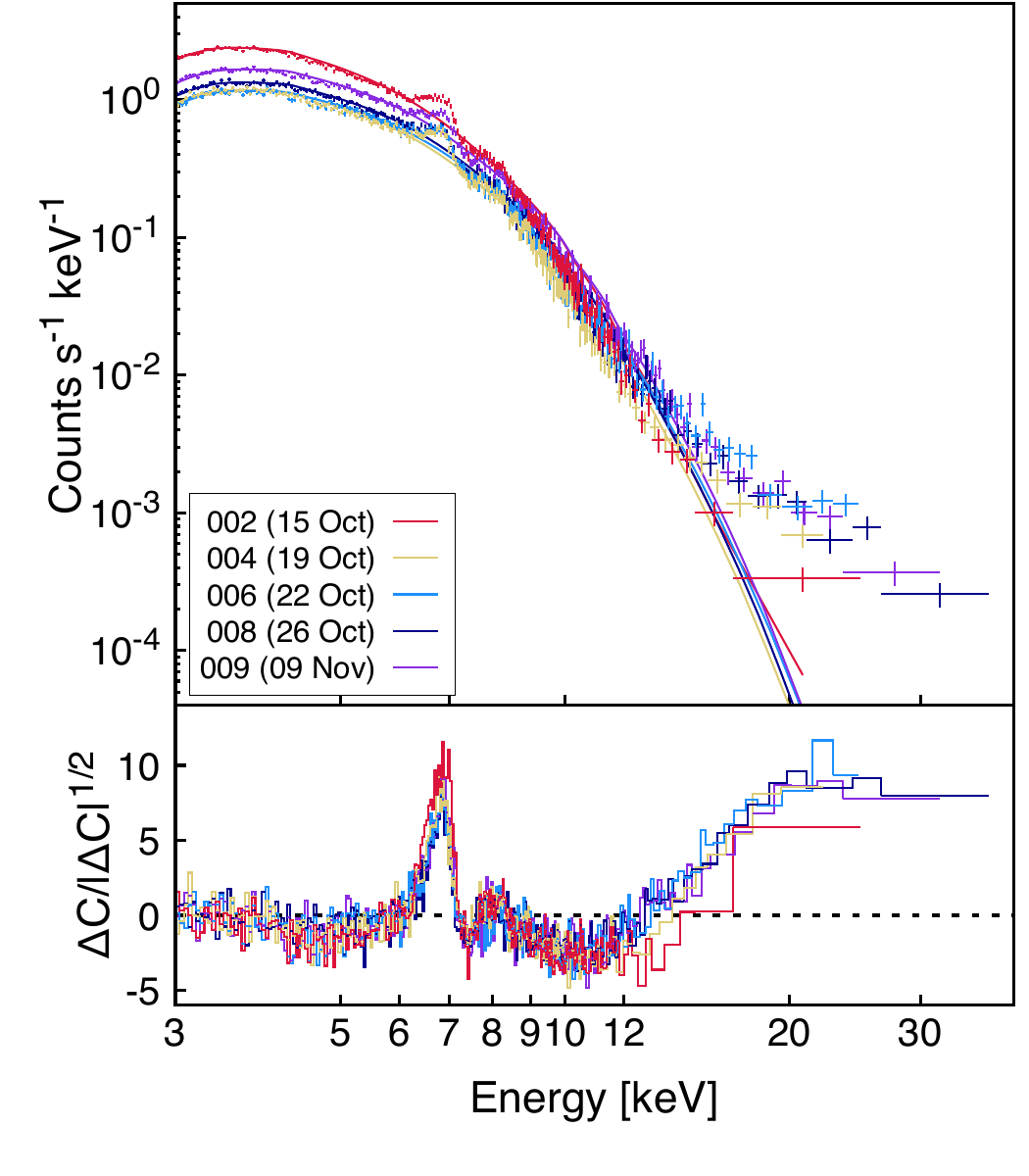}
\caption{Summary of all five \nustar\ observations (FPM A only) of \v4641\ during a month-long period in its 2021 outburst, fit with the model \texttt{TBabs*(diskbb)}. {\bf Top}: Counts data for all five observations. Overplotted lines show the model fit to each spectrum. {\bf Bottom}: Cash statistic residuals.}
\label{fig:spectrum}
\end{figure}

\begin{figure}
\includegraphics[width=\linewidth,trim={0 50 0 0}]{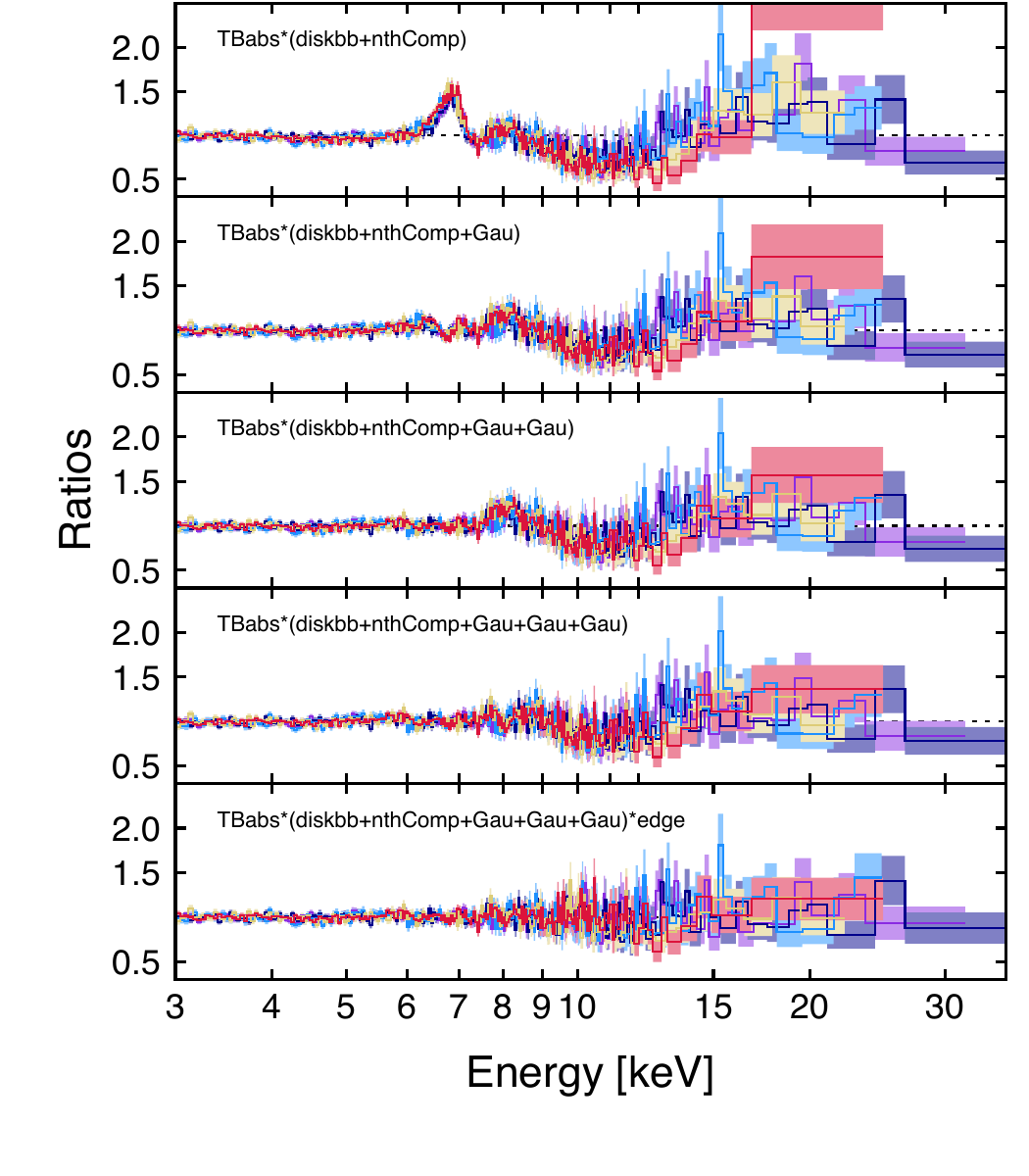}
\caption{Residuals resulting from a step-by-step progression of model fits to all five \nustar\ observations of \v4641. Each successive model contains one additional component, such that the first model is the baseline continuum consisting of absorbed disk$+$Comptonization components, followed by three additional Gaussians in turn (the $\sim6.4$~keV neutral Fe~K$\alpha$ line, the ionized Fe~K$\alpha$ line between $6.7\mbox{--}6.97$~keV, and the ionized Fe~K$\beta$ line at $\sim7.88\mbox{--}8.25$~keV), and finally an edge component at $\sim9.5$~keV.}
\label{fig:model_prog}
\end{figure}

Here we outline our full spectral analysis of the \nustar\ observations of \v4641. We performed all our spectral analysis using the Interactive Spectral Interpretation System \citep[ISIS;][]{Houck2000} version 1.6.2-51. \v4641\ was in a soft spectral state throughout the five \nustar\ observations from mid-October to mid-November 2021. Figure~\ref{fig:spectrum} shows the binned FPM A counts data for all five observations, and the statistical residuals of a simple absorbed multi-temperature disk blackbody fit to the spectra. We adopt the model 
\texttt{TBabs*(diskbb)}, where \texttt{TBabs} is the model for photoelectric absorption in the interstellar medium (we use the atomic abundances of \citealt{Wilms2000} and \citealt{Verner1996} cross sections), and \texttt{diskbb} represents a multitemperature disk blackbody \citep{Mitsuda1984,Makishima1986,Kubota1998}. We fix the hydrogen column density to $N_{\rm H}=2.5\times10^{21}~{\rm cm^{-2}}$ based on previous \nicer\ and \nustar\ X-ray spectral analysis of \v4641 \citep{Pahari2015,Shaw2022}. The residuals show clear, prominent spectral features beyond a simple disk continuum: (i) a weak power law tail at high energies; (ii) ionized emission features in the $6\mbox{--}7$~keV region, and potentially at $\sim8$~keV as well; and (iii) an apparent dip in the spectral continuum in the $8\mbox{--}12$~keV region. In addition, all these spectral features appear to be at a similar counts ratio with respect to the underlying continuum model, i.e., roughly constant equivalent width (see Figure~\ref{fig:model_prog}). The disk continuum evolves slightly over the several week long observation period: the inner disk temperature varies by $\sim10\%$ from $\sim1.3\mbox{--}1.45$~keV and disk normalization varies by a factor of $2\mbox{--}3$ from $\sim8\mbox{--}18$.  

The line features are likely associated with ionized iron K$\alpha$ and K$\beta$ emission \citep{Shaw2022}. The dip at higher energies, however, is of unknown origin. \cite{Shaw2022} note that an edge feature at $\sim9.5$~keV allows a good fit to their \nustar\ data, but do not speculate on its origin. \cite{Pahari2015} suggest that the emission line feature at $\sim8$~keV is associated with nickel, and that the $9.5$~keV feature is a strong narrow edge, possibly also nickel. Here we explore these spectral features and the underlying broadband continuum of \v4641 with step-by-step modeling. 

We start by adding a thermal Comptonization model, \texttt{nthComp} \citep{Zdziarski1996,Zycki1999}, to account for the excess high energy emission beyond $\sim10$~keV. The \texttt{nthComp} model calculates the inverse Compton up-scattering of a distribution of seed multitemperature blackbody photons from the accretion disk, with the inner disk temperature set to the value of $kT_{\rm in}$ from the \texttt{diskbb} component. The scattering particles are hot electrons originating from the coronal plasma, the geometry/size of which we leave ambiguous. The electrons follow a thermal Maxwellian with temperature $kT_{\rm e}$, which we fix to $100$~keV, given the lack of visible high energy cutoff. The fit of this absorbed disk$+$corona continuum model (\texttt{TBabs*(diskbb+nthComp)}) is shown in the top panels of Figure~\ref{fig:model_prog} as data/model ratios. The subsequent four plot panels then show our model progression. 


In the following subsections (\ref{subsec:lines} and \ref{subsec:edge}) we discuss the line and edge features in more detail. 

\vspace{-5pt}
\subsection{Fe Emission Lines}
\label{subsec:lines} 

Atop the underlying continuum we add a succession of emission line components in the form of Gaussian models (Figure~\ref{fig:model_prog}), starting with the most prominent residual feature at $\sim6.5\mbox{--}7$~keV, likely associated with the ionized Fe K$\alpha$ lines between $6.7\mbox{--}$6.97~keV. We allow a broad range for the centroid energy so as not to bias the fit, but for each spectrum its value settles close to $6.84$~keV, which is equidistant from both the ionized \ion{Fe}{25} and \ion{Fe}{26} transition energies ($6.7$ and $6.97$ keV). \nustar\ does not have sufficient energy resolution ($\sim400~{\rm eV}$ at $10$~keV) to distinguish these features. The second panel of Figure~\ref{fig:model_prog} shows this spectral fit, where it is evident that a single narrow line component is insufficient to model all the $6\mbox{--}7$~keV features. We therefore add an additional Gaussian component at $\sim6.4$~keV, once again allowing a broad range for the centroid energy when fitting. We add a final emission line component at $\sim8$~keV to account for excess residuals there, likely associated with \ion{Fe}{25} and \ion{Fe}{26} K$\beta$. We fix the line widths to $\sigma=100$~eV based on preliminary fits to each observation and the results of \cite{Shaw2022}, and to avoid blending the individual components by allowing too much freedom. We do not see evidence for strong relativistically broadened Fe K emission, in contrast with previous results focused on analysis of \nustar\ observations taken in 2014 \citep{Pahari2015}.

\begin{figure}
\includegraphics[width=\linewidth]{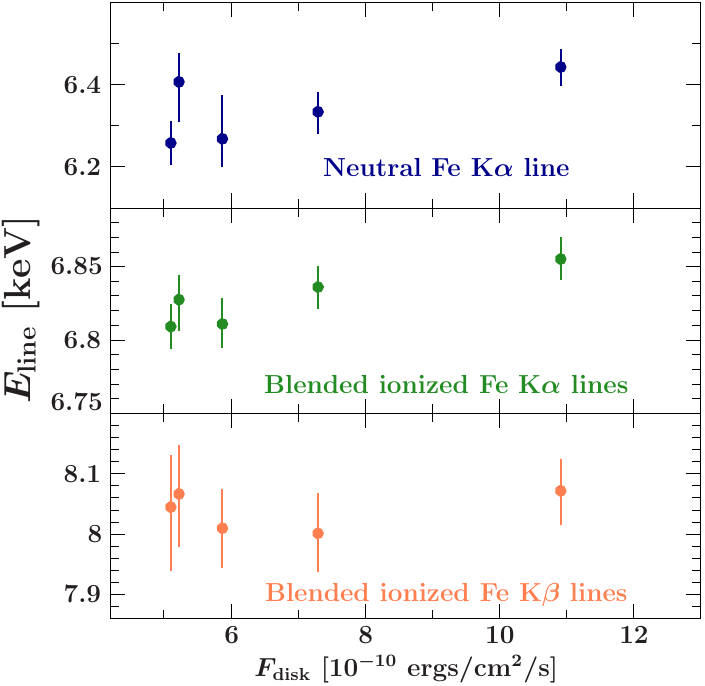}
\caption{Gaussian line energies versus disk continuum flux for all five spectra fit with model 1: \texttt{TBabs*(diskbb+nthComp+Gau+Gau+Gau)*edge}. The neutral Fe~k$\alpha$ line at $\sim6.4$~keV and the ionized Fe~k$\alpha$ lines at $\sim6.83$~keV appear to correlate with increasing disk flux.  }
\label{fig:lines}
\end{figure}

\begin{figure}
\includegraphics[width=\linewidth]{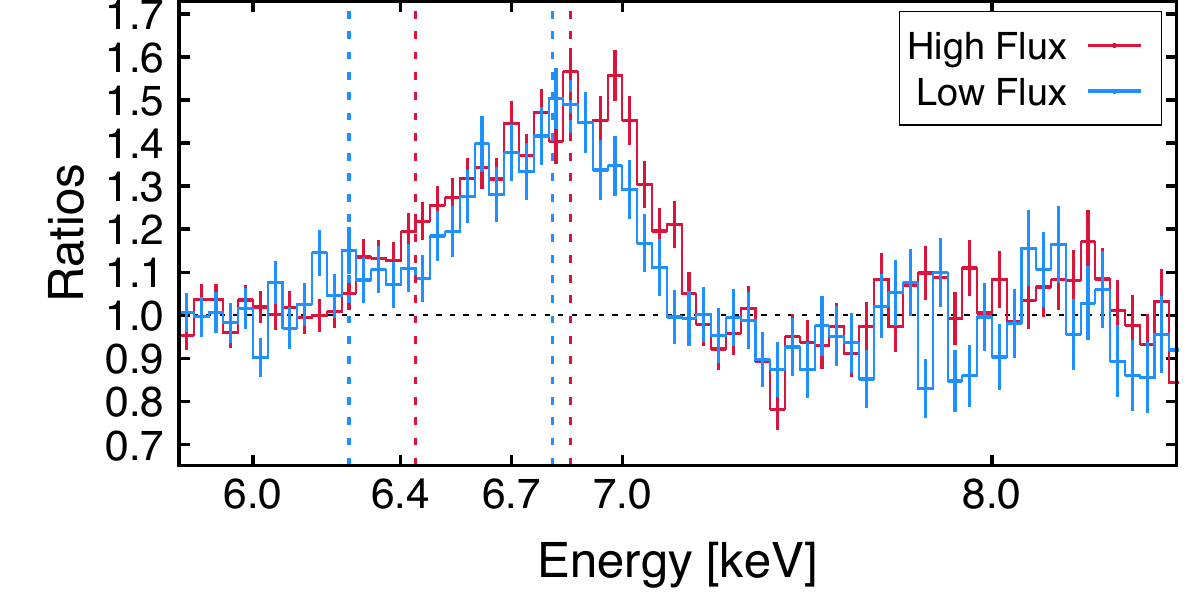}
\caption{A comparison of ratio residuals between fits of \texttt{TBabs*diskbb} to observations 1 and 3 (002 and 006), the former having roughly $2\times$ higher source flux than the latter. The dashed lines indicate the centroids of the neutral and ionized Fe~k$\alpha$ lines resulting from the full fits of model 1 to each spectrum. The higher flux spectrum has centroids at higher energies.  }
\label{fig:line_shift}
\end{figure}

Figure~\ref{fig:lines} shows the dependence of the three modeled Fe emission line energies on the disk continuum flux. Whilst the energies of all three Gaussian line components remain relatively stable, there is a mild dependence on disk flux, particularly in the case of the neutral Fe~K$\alpha$ and blended ionized Fe~K$\alpha$ lines at $\sim6.4$~keV and $\sim6.8$~keV respectively. This difference is clear when comparing the lowest and highest flux observations: the neutral Fe~K$\alpha$ component appears to shift from $\sim6.25$~keV up to $\sim6.45$~keV, while the blended ionized Fe~K$\alpha$ component shifts from $\sim6.81$~keV up to $\sim6.85$~keV. These shifts are relatively minor but, particularly in the case of the ionized component, the increase in energy may reflect an increased relative flux in the $6.97$~keV \ion{Fe}{26} component (see Figure~\ref{fig:line_shift}). If true, this would indicate an increase in the ionization of the reprocessing region. Such an increase would be expected since the disk flux differs by a factor $\sim2$ throughout the observing period, indeed the line shift is visible between observations 1 and 3 shown in Figure~\ref{fig:line_shift}, which differ in continuum flux by a factor of 2.


\subsection{Ionized Edge Feature: Testing Models}
\label{subsec:edge}

In order to unambiguously verify the edge feature (included in the fit shown in the bottom panel of Figure~\ref{fig:model_prog}), we next explore a variety of continuum models and assess the effect of such model variations on the spectral features. We chose a series of five source continuum models to perform this test, each expected to produce slight differences in the shape of either the disk continuum or Comptonization continuum, or both. The continuum constituents of the models are as follows:

\begin{itemize}
\item Model~1: \texttt{thcomp*diskbb}
\item Model~2: \texttt{thcomp*ezdiskbb}
\item Model~3: \texttt{thcomp*kerrbb}
\item Model~4: \texttt{thcomp*slimbh}
\item Model~5: \texttt{diskpn+eqpair}
\end{itemize}


\begin{figure}
\includegraphics[width=\linewidth,trim={0 120 0 0}]{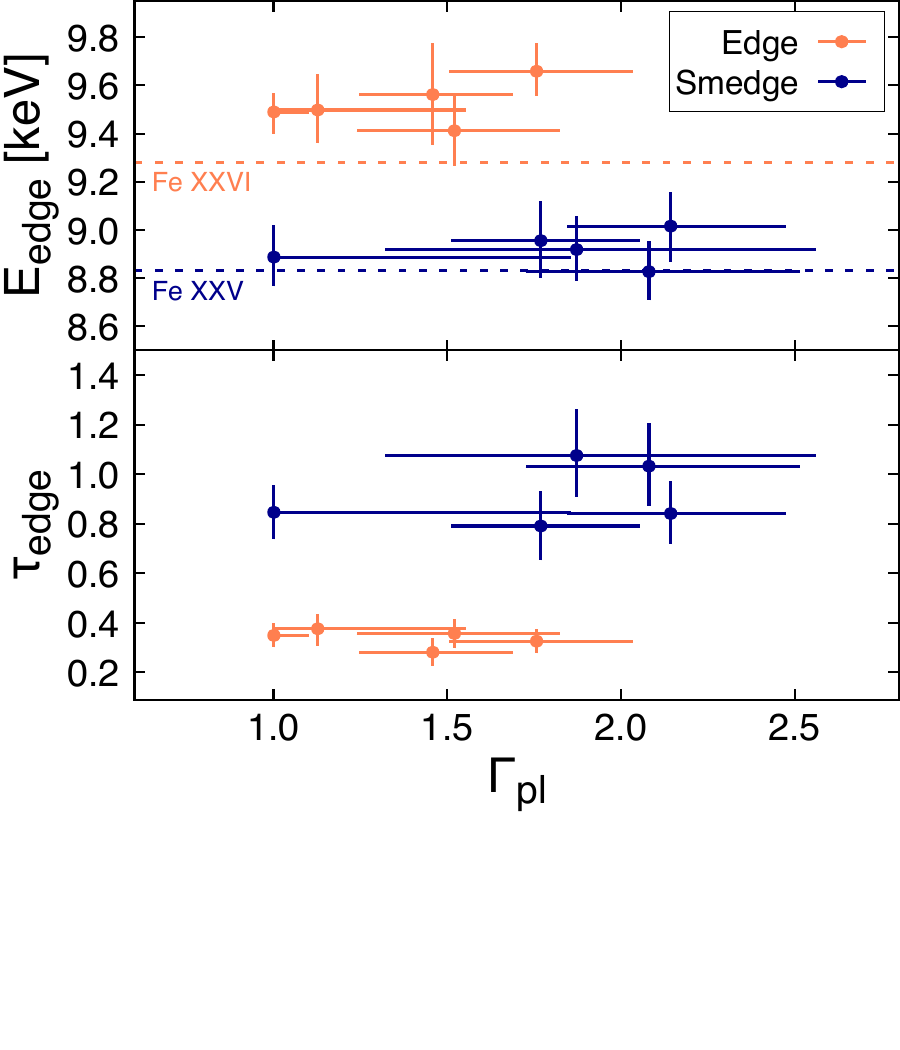}
\caption{Edge energy and optical depth as a function of the power law photon index, $\Gamma_{\rm pl}$, based on fit of Model~1a and Model~1b to all five spectra, Model~1a is identical to Model 1, Model~1b is a variation on Model 1 in which the \texttt{edge} model component is replaced by a smeared edge, or \texttt{smedge}, component. The two dashed lines in the upper panel indicate the \ion{Fe}{25} and \ion{Fe}{26} ionization levels.  }
\label{fig:edge_vs_smedge}
\end{figure}

In each case, the total model is \texttt{TBabs*(CONTINUUM MODEL+ Gaussian+Gaussian+ Gaussian) * edge}.~The variations include an accretion disk with a zero-torque boundary condition (\texttt{ezdiskbb}; \citealt{Zimmerman2005}), a general relativistic disk around a Kerr BH (\texttt{kerrbb}; \citealt{Li2005}), a relativistic slim accretion disk (\texttt{slimbh}; \citealt{Sadowski2011}), and a hybrid thermal/non-thermal Comptonization model (\texttt{eqpair}; \citealt{Coppi2000}) with a post-Newtonian disk (\texttt{diskpn}; \citealt{Gierlinski1999}). For Models $1\mbox{--}4$ we have replaced the previously adopted \texttt{nthComp} model with the convolution routine \texttt{ThComp} \citep{Zdziarski2020c}, which is an improvement on \texttt{nthComp} which treats scattering by mildy-relativistic electrons more accurately (more specifically, the regime $kT_{\rm e}\sim E_{\rm ph}$). We adopt \texttt{ThComp} because it is a convolution routine which we can apply to arbitrary input seed photon distribution, whereas \texttt{nthComp} assumes that input distribution is that of the \texttt{diskbb} model. We fit all five models to each spectrum, independently, with the edge feature switched on and then switched off.

We find that the edge feature cannot be subsumed by the combined disk and Comptonization continua. In fact, regardless of which model we apply, while the fit statistics vary slightly, the data-to-model ratios are almost identical. An edge feature is seemingly required by the data, and it appears between $9\mbox{--}10$~keV; we consistently find the edge energy prefers to settle at $\sim9.5$~keV. The edge feature reduces the Cash fit statistic by $\sim\Delta C=100$ for all Models $1\mbox{--}5$. 

Despite the lack of low-energy coverage (\nustar\ extends down to 3 keV), previous soft X-ray spectroscopic analyses have shown that a partial covering absorber may be present in \v4641 \citep{Koljonen2020,Shaw2022}. It is therefore worth verifying whether or not such an obscurer could impact the appearance of the visible edge feature, in much the same way we have tested for the impacts of alternative continuum models. Since applying Models $1\mbox{--}5$ to our data reveal no significant variations in the appearance of the edge feature, we take Model~1 and include a \texttt{TBpcf} component \citep{Wilms2000}, a commonly adopted partial covering model with two parameters, the equivalent hydrogen column ($N_{\rm H}$) and a dimensionless covering fraction ($f_{\rm cov}$). We allow both parameters to vary freely when fitting the data, but we find that the partial coverer only impacts the soft X-ray band below the Fe K region, i.e., it does not affect the veracity of the edge feature---the difference in Cash fit statistic is just $\Delta{\rm C}/\Delta{\rm DoF}=13/2$, and the edge appears identical in the ratio residuals.

To investigate the edge feature further, we took Model~1 and replaced the \texttt{edge} model component with a \texttt{smedge} component instead, in order to make a comparison of the properties of the edge in each case. The \texttt{smedge} model is a continuum edge that has a smearing kernel applied to it, representing broadening of the feature from processes such as bulk motion of the reprocessing material, or relativistic smearing \citep{Ebisawa1991}. The motivation behind applying the \texttt{smedge} model as a comparison is simply that whether the reprocessing is occurring in more distant surrounding gas, or closer to the inner accretion disk, the material is likely to be in motion \citep{Koljonen2020,Shaw2022} and consisting of different ionization zones, thus one expects a certain degree of broadening to the edge feature---whether due to motion, blended features, or scattering. We fix the smearing width to $\sigma_{\rm smedge}=2$~keV based on preliminary fits, and to avoid degeneracies with the optical depth.

\begin{figure*}
\includegraphics[width=0.5\linewidth]{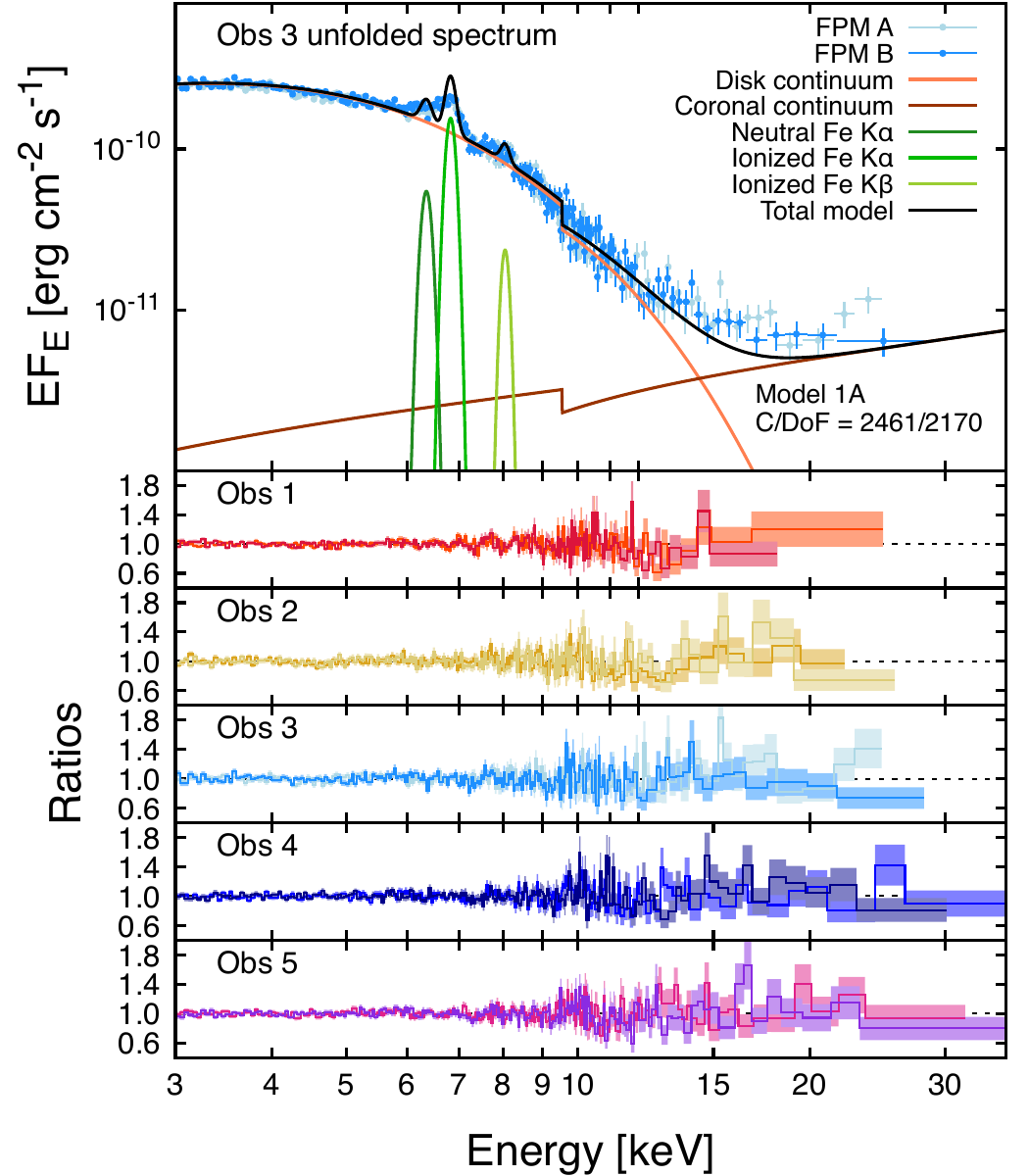}
\includegraphics[width=0.5\linewidth]{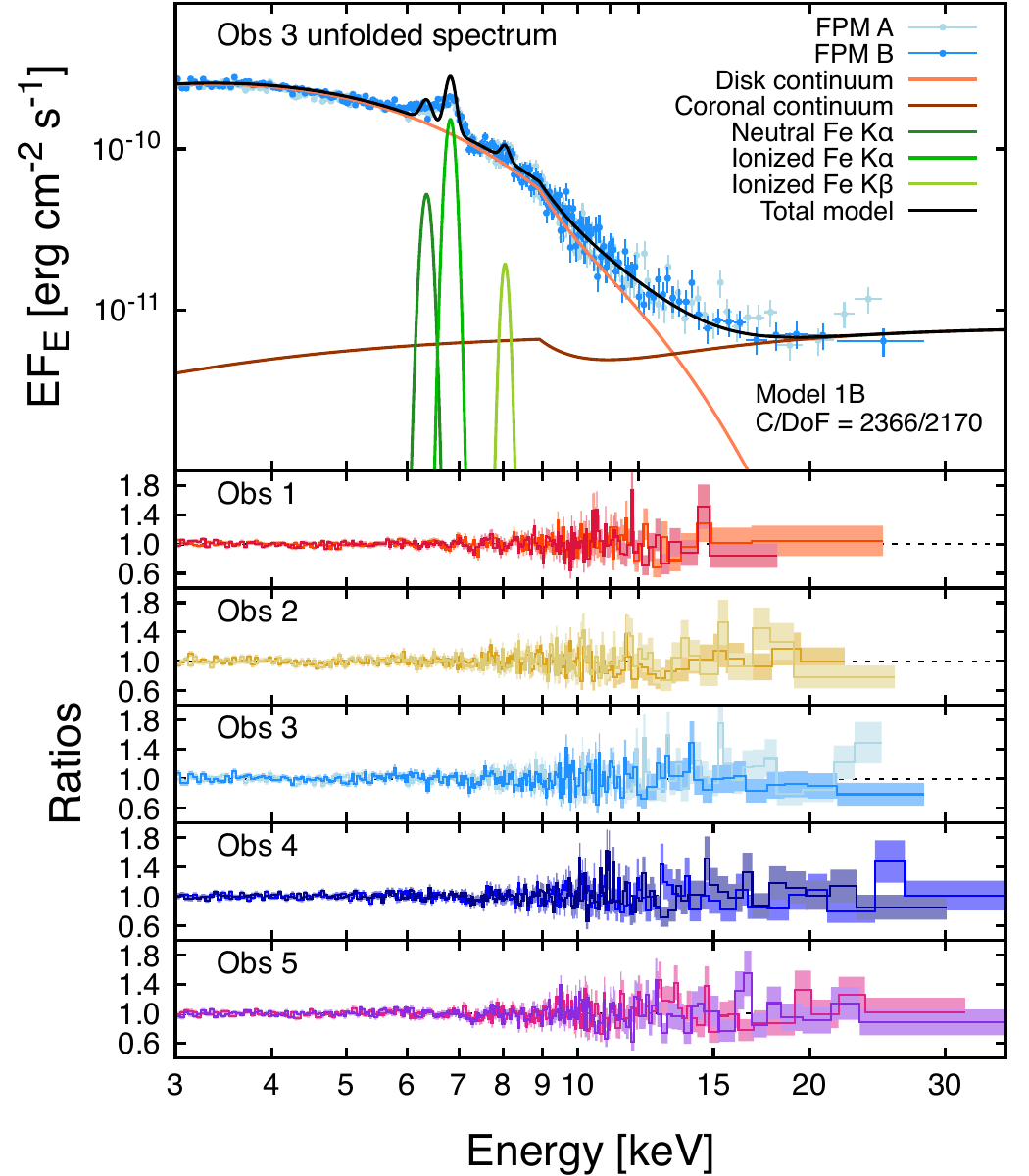}
\caption{Unfolded flux spectrum with the associated model components and ratio residuals for each fit to the five \nustar\ observations of \v4641. The left panels show fits of Model~1a, right panels show fits of Model~1b. The only difference between the two models is the treatment of the absorption edge feature around $\sim9$~keV: in Model~1a this feature is modeled using the \texttt{edge} component, in Model~1b it is instead modeled with a smeared edge or \texttt{smedge} component. }
\label{fig:final_fits}
\end{figure*}

Figure~\ref{fig:edge_vs_smedge} shows a comparison of the edge constraints (energy and optical depth) between the two model flavors 1a and 1b (\texttt{edge} vs \texttt{smedge}) as a function of the photon index of the Comptonization component, which reveals three key differences: (i) the \texttt{edge} component consistently lies at energies greater than the highest ionization continuum edge of iron, \ion{Fe}{26}, the H-like level at $9.28$~keV (the constrained value is typically closer to $\sim9.5$~keV), whereas the \texttt{smedge} component is broadly consistent with the \ion{Fe}{25} He-like level at $8.83$~keV; (ii) the optical depth of the \texttt{smedge} component is several times higher than that of the \texttt{edge} component; and (iii) the Comptonizing continuum is softer when applying the \texttt{smedge} component, and thus more consistent with what we expect in BH-LMXB soft states \citep{Remillard2006}. 

In the following Section (\ref{subsec:jointfits}) we discuss full joint spectral fits to all five \nustar\ spectra we performed based on the demonstrated relative stability of the line and edge features. 

\subsection{Joint Fits}
\label{subsec:jointfits}

Here we present full joint spectral fits to all five \nustar\ observations of \v4641 with key parameters tied between the models applied to each dataset. Based on the findings of our modeling of the individual spectra (see Sections~\ref{subsec:edge} and \ref{subsec:lines}), we selected two key models to explore in full detail with joint fits, which we name Model~1a and Model~1b:

\begin{itemize}
\item Model~1a: \texttt{TBabs*(diskbb+nthComp+Gau+Gau+Gau) * edge}
\item Model~1b: \texttt{TBabs*(diskbb+nthComp+Gau+Gau+Gau) * smedge}
\end{itemize}

where the only difference lies in the choice model component to represent the continuum edge found at $\sim9$~keV: \texttt{edge} vs \texttt{smedge}. In each case we tie the edge energy ($E_{\rm edge}$, $E_{\rm smedge}$) and optical depth ($\tau_{\rm edge}$, $\tau_{\rm smedge}$) across all fits, as well as the Gaussian line energies ($E_{\rm g1}$, $E_{\rm g2}$ and $E_{\rm g3}$)---the widths of the lines are all fixed to $\sigma=100$~eV as before, and the smeared edge width is fixed to $\sigma_{\rm smedge}=2$~keV. The remaining parameters are also all treated as before, i.e., full freedom is given to the disk ($kT_{\rm in}$ and $F_{\rm disk}$) and Comptonization ($\Gamma$ and $F_{\rm pl}$) components, which vary across all five spectra. 

Figure~\ref{fig:final_fits} shows the final fits and associated residuals, and Table~\ref{tab:params} shows the maximum likelihood estimates of the model parameters and their $90\%$ confidence intervals. We calculated confidence intervals using the \texttt{conf\_loop} parameter exploration function within ISIS, which is analogous to the \texttt{error} command typically used within \texttt{xspec}---the function loops through the free model parameters and calculates the errors, refitting the model whenever a lower fit statistic is found. 

We see the same key differences in the edge energy ($E_{\rm edge}$ vs $E_{\rm smedge}$) and optical depth ($\tau_{\rm edge}$ vs $\tau_{\rm smedge}$) between models 1a and 1b as those found in our individual spectral fits, as well as the power law continuum constraint; Model~1b gives photon indices consistent with $\Gamma\sim2$, which is typically observed in BH-LMXB soft state spectra. All other parameters are indistinguishable. Model~1b also gives an improvement in the Cash statistic of $\sim100$ over Model~1a (2366 vs 2461) for the same degrees of freedom. All these factors confirm that the observed absorption edge feature at $\sim9$~keV is likely a smeared continuum edge originating from \ion{Fe}{25} at $8.83$~keV, i.e., we are seeing absorption of the continuum emission via highly ionized Fe in a photo-ionized gas surrounding the accretion flow which is in motion.


\def\MAallobsCash{$2461/2170$}
\def\MAallobsNh{$0.25^{a}$}
\def\MAallobsEgauOne{$6.35\pm0.03$}
\def\MAallobsEgauTwo{$6.829\pm0.007$}
\def\MAallobsEgauThree{$8.04\pm0.03$}
\def\MAallobsEedge{$9.54^{+0.05}_{-0.06}$}
\def\MAallobstauedge{$0.33\pm0.02$}

\def\MAkTinOne{$1.263\pm0.005$}
\def\MAGammaOne{$<1.09$}
\def\MAFgOneOne{$2.9\pm0.4$}
\def\MAFgTwoOne{$9.7\pm0.4$}
\def\MAFgThreeOne{$1.4\pm0.2$}
\def\MAfdiskOne{$10.92\pm0.07$}
\def\MAfplOne{$0.5^{+0.1}_{-0.4}$}

\def\MAkTinTwo{$1.332\pm0.008$}
\def\MAGammaTwo{$<1.4$}
\def\MAFgOneTwo{$1.6\pm0.3$}
\def\MAFgTwoTwo{$5.3\pm0.3$}
\def\MAFgThreeTwo{$0.7\pm0.2$}
\def\MAfdiskTwo{$5.24\pm0.04$}
\def\MAfplTwo{$0.3^{+0.8}_{-0.2}$}

\def\MAkTinThree{$1.38\pm0.01$}
\def\MAGammaThree{$1.5\pm0.3$}
\def\MAFgOneThree{$2.1\pm0.3$}
\def\MAFgTwoThree{$5.1\pm0.3$}
\def\MAFgThreeThree{$0.6\pm0.2$}
\def\MAfdiskThree{$5.12\pm0.04$}
\def\MAfplThree{$0.31^{+0.12}_{-0.05}$}

\def\MAkTinFour{$1.345^{+0.009}_{-0.008}$}
\def\MAGammaFour{$1.5\pm0.2$}
\def\MAFgOneFour{$1.3\pm0.3$}
\def\MAFgTwoFour{$4.6\pm0.3$}
\def\MAFgThreeFour{$0.6\pm0.2$}
\def\MAfdiskFour{$5.86\pm0.04$}
\def\MAfplFour{$0.24^{+0.05}_{-0.03}$}

\def\MAkTinFive{$1.366\pm0.008$}
\def\MAGammaFive{$1.7^{+0.3}_{-0.2}$}
\def\MAFgOneFive{$2.0\pm0.3$}
\def\MAFgTwoFive{$5.8\pm0.3$}
\def\MAFgThreeFive{$0.8\pm0.2$}
\def\MAfdiskFive{$7.29\pm0.05$}
\def\MAfplFive{$0.23^{+0.03}_{-0.02}$}
  

\def\MBallobsCash{$2366/2170$}
\def\MBallobsNh{$0.25^{a}$}
\def\MBallobsEgauOne{$6.35\pm0.03$}
\def\MBallobsEgauTwo{$6.827\pm0.007$}
\def\MBallobsEgauThree{$8.04^{+0.03}_{-0.04}$}
\def\MBallobsEedge{$8.91\pm0.06$}
\def\MBallobstauedge{$0.89\pm0.06$}

\def\MBkTinOne{$1.273^{+0.005}_{-0.006}$}
\def\MBGammaOne{$<2$}
\def\MBFgOneOne{$2.7\pm0.4$}
\def\MBFgTwoOne{$9.5\pm0.4$}
\def\MBFgThreeOne{$1.2\pm0.2$}
\def\MBfdiskOne{$10.84\pm0.07$}
\def\MBfplOne{$0.6^{+0.2}_{-0.5}$}

\def\MBkTinTwo{$1.33\pm0.01$}
\def\MBGammaTwo{$1.6^{+0.5}_{-0.4}$}
\def\MBFgOneTwo{$1.5\pm0.3$}
\def\MBFgTwoTwo{$5.3\pm0.3$}
\def\MBFgThreeTwo{$0.7\pm0.2$}
\def\MBfdiskTwo{$5.20^{+0.04}_{-0.05}$}
\def\MBfplTwo{$0.18^{+0.13}_{-0.03}$}

\def\MBkTinThree{$1.37\pm0.01$}
\def\MBGammaThree{$1.9\pm0.3$}
\def\MBFgOneThree{$2.0\pm0.3$}
\def\MBFgTwoThree{$5.0\pm0.3$}
\def\MBFgThreeThree{$0.5\pm0.2$}
\def\MBfdiskThree{$5.06^{+0.05}_{-0.06}$}
\def\MBfplThree{$0.31^{+0.03}_{-0.02}$}

\def\MBkTinFour{$1.35\pm0.01$}
\def\MBGammaFour{$1.9^{+0.3}_{-0.2}$}
\def\MBFgOneFour{$1.2\pm0.3$}
\def\MBFgTwoFour{$4.5\pm0.3$}
\def\MBFgThreeFour{$0.5\pm0.2$}
\def\MBfdiskFour{$5.80\pm0.05$}
\def\MBfplFour{$0.26\pm0.02$}

\def\MBkTinFive{$1.37\pm0.01$}
\def\MBGammaFive{$2.2\pm0.3$}
\def\MBFgOneFive{$1.9^{+0.4}_{-0.3}$}
\def\MBFgTwoFive{$5.7\pm0.3$}
\def\MBFgThreeFive{$0.7\pm0.2$}
\def\MBfdiskFive{$7.19^{+0.06}_{-0.07}$}
\def\MBfplFive{$0.29^{+0.06}_{-0.04}$}

\begin{deluxetable*}{lcccccccccc}
\tablecaption{Maximum likelihood estimates and 90\% confidence intervals of all parameters in spectral fitting the five \nustar\ observations of \v4641 with Model 1a (\texttt{TBabs*(diskbb+nthComp+Gau+Gau+Gau)*edge}) and Model 1b (\texttt{TBabs*(diskbb+nthComp+Gau+Gau+Gau)*smedge})}.\label{tab:params}
\tablecolumns{6}
\tablehead{
\colhead{Parameters} & 
\multicolumn{5}{c}{Model 1a}
}
\startdata
\hline
C/DoF & & & \MAallobsCash\ & & \\
$N_{\rm H}~[10^{22}~{\rm cm}^{-2}]$ & & & \MAallobsNh\ & & \\
$E_{\rm g1}$~[keV] & & & \MAallobsEgauOne\ & & \\
$E_{\rm g2}$~[keV] & & & \MAallobsEgauTwo\ & & \\
$E_{\rm g3}$~[keV] & & & \MAallobsEgauThree\ & & \\
$E_{\rm edge}$~[keV] & & & \MAallobsEedge\ & & \\
$\tau_{\rm edge}$ & & & \MAallobstauedge\ & &  \\
\hline
& \colhead{Obs 1} & \colhead{Obs 2} & \colhead{Obs 3} & \colhead{Obs 4} & \colhead{Obs 5} \\
\hline
$kT_{\rm in}$~[keV] & \MAkTinOne\ & \MAkTinTwo\ & \MAkTinThree\ & \MAkTinFour\ & \MAkTinFive\ \\
$\Gamma$ & \MAGammaOne\ & \MAGammaTwo\ & \MAGammaThree\ & \MAGammaFour\ & \MAGammaFive\ \\
$F_{\rm g1}~{\rm [10^{-4}~Ph~cm^{-2}~s^{-1}]}$ & \MAFgOneOne\ & \MAFgOneTwo\ & \MAFgOneThree\ & \MAFgOneFour\ & \MAFgOneFive\ \\
$F_{\rm g2}~{\rm [10^{-4}~Ph~cm^{-2}~s^{-1}]}$ & \MAFgTwoOne\ & \MAFgTwoTwo\ & \MAFgTwoThree\ & \MAFgTwoFour\ & \MAFgTwoFive\ \\
$F_{\rm g3}~{\rm [10^{-4}~Ph~cm^{-2}~s^{-1}]}$ & \MAFgThreeOne\ & \MAFgThreeTwo\ & \MAFgThreeThree\ & \MAFgThreeFour\ & \MAFgThreeFive\ \\
$F_{\rm disk}~{\rm [10^{-10}~erg~cm^{-2}~s^{-1}]}$ & \MAfdiskOne\ & \MAfdiskTwo\ & \MAfdiskThree\ & \MAfdiskFour\ & \MAfdiskFive\ \\
$F_{\rm pl}~{\rm [10^{-10}~erg~cm^{-2}~s^{-1}]}$ & \MAfplOne\ & \MAfplTwo\ & \MAfplThree\ & \MAfplFour\ & \MAfplFive\ \\
\hline
\hline
\multicolumn{1}{c}{Parameters} & \multicolumn{5}{c}{Model 1b} \\
\hline
C/DoF & & & \MBallobsCash\ & & \\
$N_{\rm H}~[10^{22}~{\rm cm}^{-2}]$ & & & \MBallobsNh\ & & \\
$E_{\rm g1}$~[keV] & & & \MBallobsEgauOne\ & & \\
$E_{\rm g2}$~[keV] & & & \MBallobsEgauTwo\ & & \\
$E_{\rm g3}$~[keV] & & & \MBallobsEgauThree\ & & \\
$E_{\rm smedge}$~[keV] & & & \MBallobsEedge\ & & \\
$\tau_{\rm smedge}$ & & & \MBallobstauedge\ & &  \\
\hline
& \colhead{Obs 1} & \colhead{Obs 2} & \colhead{Obs 3} & \colhead{Obs 4} & \colhead{Obs 5} \\
\hline
$kT_{\rm in}$~[keV] & \MBkTinOne\ & \MBkTinTwo\ & \MBkTinThree\ & \MBkTinFour\ & \MBkTinFive\ \\
$\Gamma$ & \MBGammaOne\ & \MBGammaTwo\ & \MBGammaThree\ & \MBGammaFour\ & \MBGammaFive\ \\
$F_{\rm g1}~{\rm [10^{-4}~Ph~cm^{-2}~s^{-1}]}$ & \MBFgOneOne\ & \MBFgOneTwo\ & \MBFgOneThree\ & \MBFgOneFour\ & \MBFgOneFive\ \\
$F_{\rm g2}~{\rm [10^{-4}~Ph~cm^{-2}~s^{-1}]}$ & \MBFgTwoOne\ & \MBFgTwoTwo\ & \MBFgTwoThree\ & \MBFgTwoFour\ & \MBFgTwoFive\ \\
$F_{\rm g3}~{\rm [10^{-4}~Ph~cm^{-2}~s^{-1}]}$ & \MBFgThreeOne\ & \MBFgThreeTwo\ & \MBFgThreeThree\ & \MBFgThreeFour\ & \MBFgThreeFive\ \\
$F_{\rm disk}~{\rm [10^{-10}~erg~cm^{-2}~s^{-1}]}$ & \MBfdiskOne\ & \MBfdiskTwo\ & \MBfdiskThree\ & \MBfdiskFour\ & \MBfdiskFive\ \\
$F_{\rm pl}~{\rm [10^{-10}~erg~cm^{-2}~s^{-1}]}$ & \MBfplOne\ & \MBfplTwo\ & \MBfplThree\ & \MBfplFour\ & \MBfplFive\ \\
\enddata
\tablecomments{ $^a$Frozen parameter. \\
In all fits the following parameters are frozen: $N_{\rm H}=2.5\times10^{21}~{\rm cm^{-2}}$, $kT_{\rm e}=100$~keV, $\sigma_{\rm g}=100$~eV (Gaussian line widths), $\sigma_{\rm smedge}=2$~keV (the width of the smeared edge as adopted in Model 1b). In addition, we calculate the unabsorbed line and continuum fluxes using the \texttt{cpflux} and \texttt{cflux} convolution kernels respectively. We fix the integration limits of the line photon fluxes to $E_{\rm min}=0.5$~keV, $E_{\rm max}=10$~keV. The \texttt{diskbb} flux is calculated over the range $0.01\mbox{--}50$~keV, and the \texttt{nthComp} flux over the range $0.01\mbox{--}200$~keV.} 
\end{deluxetable*}

\begin{figure}
\includegraphics[width=\linewidth,trim={0 120 0 0}]{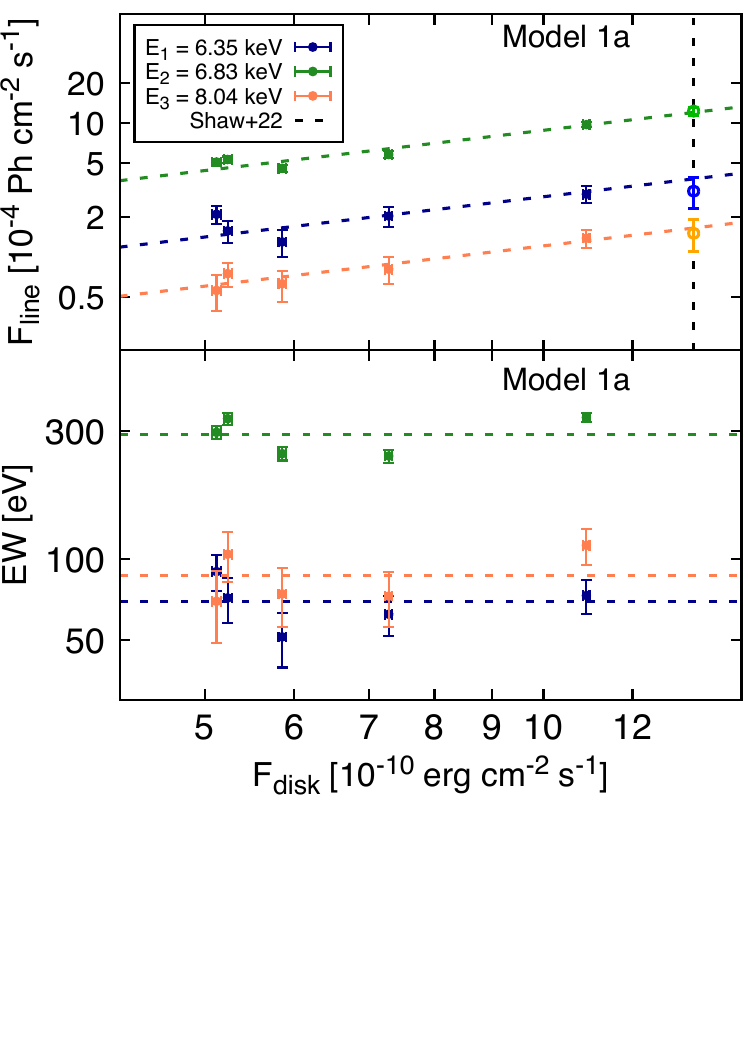}
\caption{{\bf Top}: Emission line fluxes versus the disk continuum flux for Model 1a fits. The dotted lines show fits to the trend of each emission line energy vs disk flux with a slope of 1, demonstrating a one-to-one correlation between the line and continuum fluxes. The lighter points coincident with the vertical black line show the same constraints found by \cite{Shaw2022} resulting from similar fits to earlier \nustar\ observations of \v4641. Note these points are overplotted and not part of the regression. {\bf Bottom}:  Line equivalent width (EW) vs disk continuum flux for all three modeled Gaussian emission lines. Dotted lines show the flat mean EW value fit to the constraints.}
\label{fig:line_fluxes}
\end{figure}

\begin{figure}
\includegraphics[width=\linewidth,trim={0 0 20 0}]{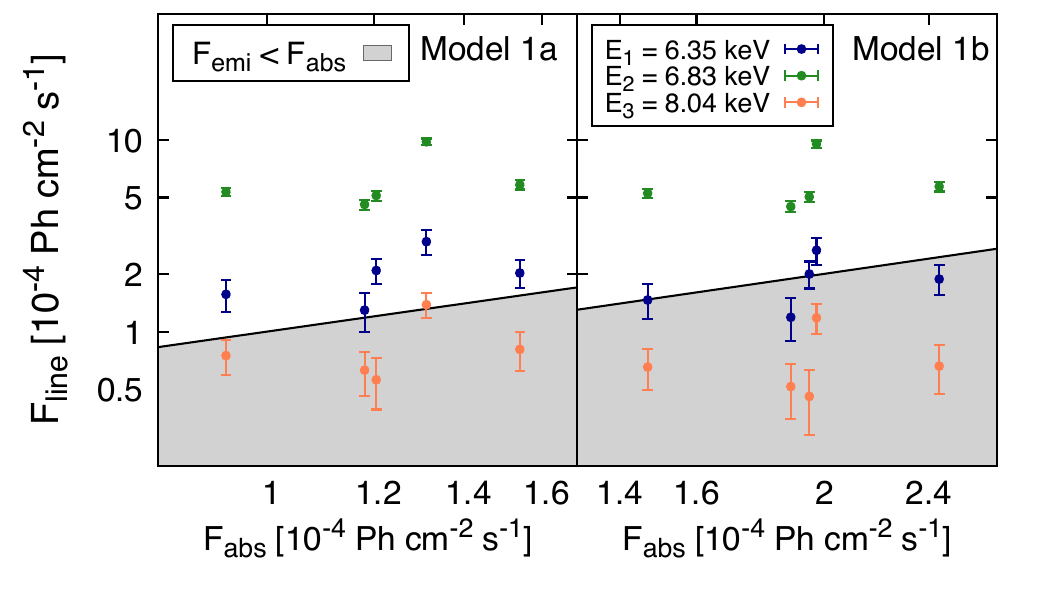}
\caption{Emission line fluxes versus the absorbed photon flux, via the ionized iron continuum edge, showing constraints from both Model 1a (left) and Model 1b (right) fits. The shaded zone shows the region in which the emitted photon flux is less than the absorbed photon flux, which violates photon conservation \citep{Reynolds2009}. }
\label{fig:fline_v_fedge}
\end{figure}



\vspace{-20pt}
\subsubsection{Parameter trends}
\label{subsec:trends}

Figure~\ref{fig:line_fluxes} shows the emitted photon flux (top) and equivalent width (EW; bottom) for each modeled Gaussian line (at $\sim6.35$~keV, $6.83$~keV, and $8.04$~keV respectively) against the disk continuum flux, resulting from Model 1a fits (we choose to show only Model 1a fit results here since the trends are very similar for both Model 1a and 1b). There is a distinct positive correlation between the flux of each line and the disk continuum flux, which is confirmed by a regression line fit to each trend respectively. Interestingly, the EWs are also quite stable, showing no correlation with disk flux, though there are some variations around the mean value. This shows there is likely a connection between the observed disk continuum emission and the reprocessed line emission, i.e., we must be seeing at least some portion of what the photoionized gas sees. 

Table~\ref{tab:params} shows that the power law flux remains low compared to the disk flux ($F_{\rm pl} \sim 0.05 F_{\rm disk}$), and statistically consistent with being unchanged between observations. The power law photon index ($\Gamma$) shows some variation but is poorly constrained in all our fits---it is always consistent with a flat spectrum, $\Gamma=2$. 

Figure~\ref{fig:fline_v_fedge} shows the line photon fluxes against the photon flux absorbed in the edge component. We compare constraints from Models 1a (left) and 1b (right), since the \texttt{edge} and \texttt{smedge} components give different absorbed fluxes. The shaded regions indicate the phase space in which photon conservation is violated, i.e, where the number of photons emitted is less than the number absorbed by the gas ($F_{\rm emi} < F_{\rm abs}$), which is unphysical if we assume the line emission is re-emission from the absorbing gas \citep{Reynolds2009}. This is outlined in detail in \cite{Reynolds2009}, but the principle argument is as follows: for every photon absorbed there must be a corresponding H-like photon liberated from the Fe atom. This corresponds to the \ion{Fe}{26} K$\alpha$ and K$\beta$ lines at $6.97$~keV and $8.25$~keV respectively. Since we have modeled the H-like and He-like lines as blends (both the Fe K$\alpha$ and Fe K$\beta$ regions), we cannot verify precisely whether enough flux is being re-emitted. However, it seems likely that $F_{\rm emi}>F_{\rm abs}$ based upon the flux of the $6.83$~keV line, which is significantly greater than $F_{\rm abs}$ for both Model 1a and Model 1b, i.e., regardless of whether the absorption edge is narrow or smeared. This is physically consistent with at least some of the ionized photons originating from the same gaseous region that is absorbing the disk emission. 

\section{Discussion and Conclusions}
\label{sec:discussion}

We have performed a detailed spectral modeling analysis of five \nustar\ observations of \v4641\ during its 2021 outburst. The observations were spread over a month-long period, and we track the changes to its continuum and line features over this period. Our key findings are as follows:
\begin{itemize}
\item {\it The broadband X-ray spectrum of \v4641\ in the soft states consists of a dominant bright multitemperature disk and a weak Comptonized component, with several ionized Fe K lines and an ionized Fe K edge component.}
\item {\it The Fe K line fluxes correlate linearly with the disk continuum flux, and are quite stable in energy, consistent with the baseline near-neutral $6.4$~keV Fe K$\alpha$ line, the ionized $6.7\mbox{--}7$~keV Fe K$\alpha$ lines, and the ionized $\sim8$~keV Fe K$\beta$ lines, modeled as three blended components. }
\item {\it We see an absorption edge component that is stable in energy and optical depth, interpreted as a highly ionized Fe K edge, which likely has some degree of smearing.}
\end{itemize}

In the following subsections we discuss the implications of our spectral fitting results and offer further interpretations.

\subsection{Reflection?}
\label{subsec:reflection}

\begin{figure}
\includegraphics[width=\linewidth]{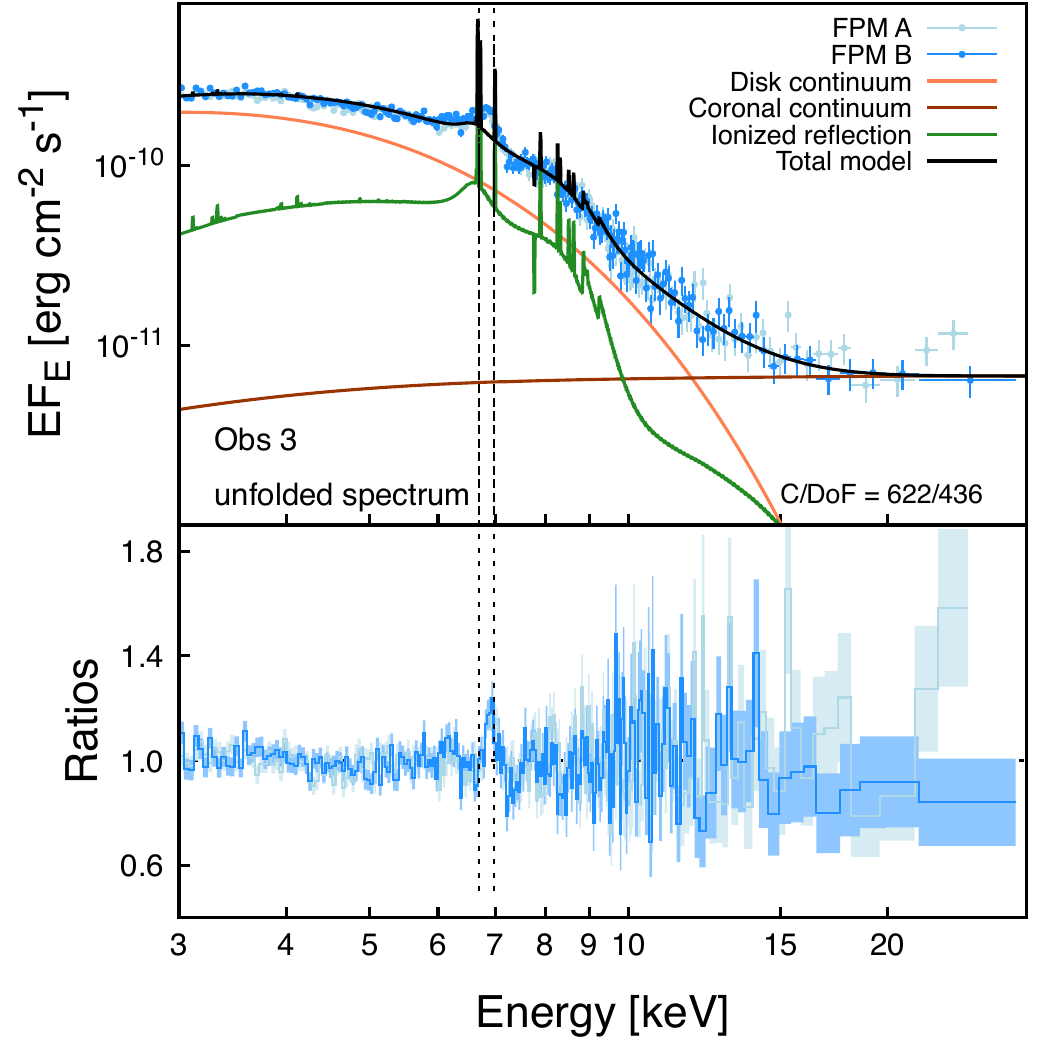}
\caption{The unfolded \nustar\ spectrum and ratio residuals resulting from observation 3 (006) of \v4641\ fit with the model \texttt{TBabs*(diskbb+nthComp+xillverNS)}. The \texttt{xillverNS} component is an ionized reflection model which assumes an irradiating blackbody spectrum atop an infinite slab (accretion disk), given by a single-temperature blackbody continuum. }
\label{fig:xillverns}
\end{figure}

Given \v4641\ is in the soft spectral state, and there are clear signs of strong reprocessing in the form of ionized Fe K emission lines and an ionized Fe K edge with $\tau\sim1$, it is worth asking the question, could the observed emission and absorption (edge) features simply be due to reflection off the accretion disk? Furthermore, and perhaps more pertinently, are we looking at just a single reprocessing region, or multiple \citep{Shaw2022}? 


Figure~\ref{fig:xillverns} shows a fit of the model \texttt{TBabs*(diskbb+nthComp+xillverNS)} to \nustar\ observation 3 (006) of \v4641. The \texttt{xillverNS} component is a physically-motivated model, calculating reflection of an irradiating blackbody off an ionized slab \citep{Garcia2022}. Our aim is to see if it can account for the three Gaussian line emission components and the continuum absorption edge at $\sim9\mbox{--}10$~keV. We set the blackbody temperature, $kT_{\rm bb}$, equal to the inner disk temperature, $kT_{\rm in}$, and allow the ionization, $\log\xi$, iron abundance, $A_{\rm Fe}$, and inclination, $i$, of the reflector to vary freely. 

There are a few key things to notice about this model fit (Figure~\ref{fig:xillverns}). First, the reflection model adequately accounts for the smeared K edge. Second, the combined disk, power law, and reflection model fits well with the general phenomenological interpretation of a bright disk and a roughly flat, faint power law spectrum (just as is the case when we apply a smeared edge, the broadened dip inherent to the reflection component allows for a softer Comtponization component, as opposed to when we model the absorption feature as a sharp edge). Finally, and most interestingly, the reflection component cannot model all of the ionized line features. The fit accounts well for the near-neutral Fe K$\alpha$ emission at $6.4$~keV and the ionized Fe K$\beta$ emission at $\sim8$~keV, but it fails to account for the strong ionized Fe K$\alpha$ line at $6.7\mbox{--}7$~keV. We also tried including an additional reflected Comptonization component, {\tt xillverCp} \citep{Garcia2014}, thus adopting two reflection components, {\tt xillverNS+xillverCp}. We found the observed power law flux is not sufficient to provide enough reflected flux to contribute significantly to the emission line features. This is further confirmation that we are quite likely viewing more than one photoionized region, or photoionized gas that has variable ionization---i.e. a single-zone photoionization model cannot account for all the emission lines simultaneously. The {\tt relxillNS} model, i.e., the relativistic disk reflection version of {\tt xillverNS}, does not yet contain the utility of an ionization gradient with disk radius \citep{Garcia2022}. Such a model variant may be preferable to apply to these data, in order to reproduce emission lines from regions of differing ionization. 

 \cite{Shaw2022} also proposed multiple reprocessing regions to explain their \chandra\ grating spectra. They argued that the ionized Fe K emission likely originates from a single region, with the longer wavelength line species (from e.g., S/Ar, Si, Mg, and Ne) originating from a separate region. Unfortunately, they did not detect the $6.4$~keV Fe~K$\alpha$ line with enough statistical significance to model it with XSTAR. We suggest that in fact there are multiple Fe K emission regions, or at least, the reprocessing gas has a stratified ionization, but we cannot rule out the possibility that a more accurate continuum model may allow for improved modeling of the ionized lines. We discuss the details of the continuum emission in the following subsection. 

\subsection{Disk Continuum and Inner Radius}

\begin{figure}
\includegraphics[width=\linewidth]{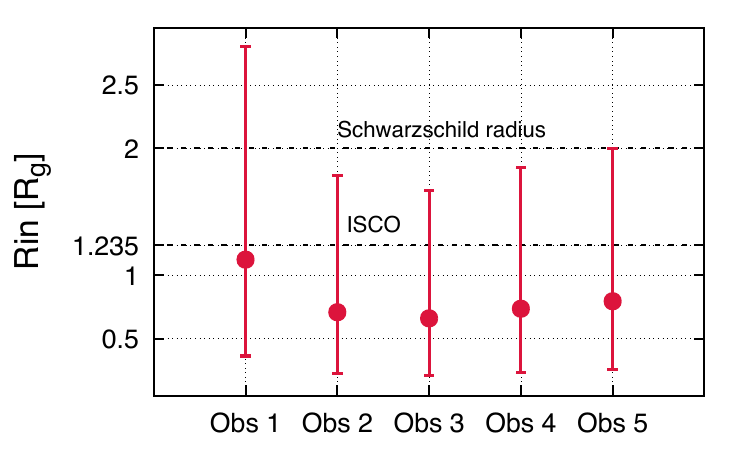}
\caption{Constraints on the inner radius of the accretion disk resulting from joint fits of Model~1b to all five \nustar\ observations of \v4641 (derived from the {\tt diskbb} component). The error bars reflect the full range of possible $R_{\rm in}$ values based on the uncertainties on fit parameters ($F_{\rm disk}$ and $T_{\rm in}$), binary characteristics ($D$, $M_{\rm BH}$, and $i$), and color correction $f_{\rm col}$.} 
\label{fig:rin}
\end{figure}

The apparent inner radius of the observed accretion disk can be deduced from the measured disk flux ($F_{\rm disk}$) and temperature ($T_{\rm in}$)---via the modeled {\tt diskbb} component---through the following relation:
\begin{equation}
\label{eq:rin}
r_{\rm in} = \left(\frac{F_{\rm disk} D^2}{2 (\cos{i}) \sigma T_{\rm in}^4}\right)^{1/2}\,\,\, ,
\end{equation}
where $D=6.2\pm0.7$~kpc is the distance to the source, and $i=72\pm4$~deg is the disk inclination. The true disk inner radius, $R_{\rm in}$, then depends on the proximity of the disk to the BH horizon (due to relativistic effects) and the degree of scattering in the disk atmosphere; the latter is primarily responsible for the decoupling of color temperature and effective temperature, which is parameterized by a color correction such that such that $R_{\rm in}=f_{\rm col} \cdot r_{\rm in}$, with typical values of roughly $f_{\rm col}\sim1.5\mbox{--}2$. The uncertainty on the inner disk radius can be impacted by the accuracy of the color correction factor. An additional caveat is that the {\tt diskbb} model has an unphysical inner-boundary condition (it assumes a non-zero torque at the inner edge; \citealt{Kubota1998,Zimmerman2005}), which we do not account for here. 

Figure~\ref{fig:rin} shows the $R_{\rm in}$ constraints resulting from the disk continuum fits to all five \nustar\ observations of \v4641. We use Equation~\ref{eq:rin} to calculate apparent inner radii, fold in the uncertainty in color correction ($f_{col}=1.5\pm0.5$), and convert to $R_{\rm g}=GM_{\rm BH}/c^2$, adopting the full range of possible values of the fit parameters ($F_{\rm disk}$, $T_{\rm in}$) and system characteristics ($D$, $i$, $M_{\rm BH}$), as opposed to propagating errors in quadrature, thus exploring the full uncertainty landscape. Most values lie within the Schwarzschild radius of the BH ($R_{\rm S}=2~R_{\rm g}$), assuming a BH mass of $M_{\rm BH}=6.4~M_{\odot}$, and all have valid (even preferred) solutions within the innermost stable circular orbit (ISCO) of a maximally spinning ($a_{\star}=0.998$) BH. Most importantly, our results are consistent with the inner disk being outside the BH horizon and the ISCO. 

We further tested the veracity of our inner radius constraints by replacing the {\tt diskbb} component with a more physical disk continuum, {\tt kerrbb}, which treats the relativistic effects to completion. If the fit results of Models 1a and 1b are valid, we should expect to find that {\tt kerrbb} fits retrieve preferably higher spin values---this is indeed what we find. An exploration of the spin parameter space---via a {\tt steppar} run of 50 steps in the range $0\le a_{\star} \le 1$---results in contours of increasing likelihood from low to high spins, i.e., maximal spin is strongly preferred. 


A number of authors have inferred obscuration of the central engine based on the soft continuum of \v4641\ in its many outbursts (e.g., \citealt{Revnivtsev2002,Maitra2006,Koljonen2020,Shaw2022}), which appears faint enough such that the disk is unfeasibly small (i.e., within the BH horizon). The extent to which the continuum is obscured during the 2021 outburst, however, may be less extreme, as can be seen in Figure~\ref{fig:rin}. We instead find solutions that are feasible, i.e., consistent with being outside the BH horizon and beyond the ISCO of a maximally spinning BH, despite the constraints skewing to low values. Since we are comparing results from observations of \v4641\ taken at various times across different outbursts, the contrasts may simply be due to intrinsic variations in the disk component. Thus, whilst we do see a hint of obscuration in the low intrinsic disk luminosity ($L_{\rm X}\sim0.003\mbox{--}0.006~L_{\rm Edd}$) and associated disk radius, the \nustar\ data alone do not allow us to disentangle a brighter, obscured component---i.e., via a partial covering model, as applied by \cite{Shaw2022} to the \chandra-HETG data---and an intrinsically faint disk.



\subsection{Ionized Fe K edge: Column of the Photoionized Gas}

We have shown in Sections~\ref{subsec:edge} and \ref{subsec:jointfits} that the 2021 \nustar\ spectra of \v4641\ are best fit with the inclusion of a smeared edge component to model the absorption edge feature present in the data. This edge feature has been detected and interpreted before \citep{Pahari2015,Shaw2022}. \cite{Pahari2015} reported a strong, narrow edge feature at $\sim9.5$~keV, and argued that it is a sharp Ni feature associated with a Ni emission line at $\sim8$~keV. However, in the 2021 outburst, the energy of the edge aligns more closely with the K edges of highly-ionized Fe than Ni. Since our analysis shows that a smeared edge at $\sim8.9$~keV is preferred (Model~1b), we interpret the feature as the \ion{Fe}{25} $8.83$~keV continuum edge. It is worth noting that the degree of smearing applied to the edge, with a width of $2$~keV at edge energy $\sim9$~keV, implies a mildly relativistic Doppler shift. Since we do not see this degree of Doppler shifting to the observed emission lines, the smearing may be caused by photon scattering, rather than Doppler motions. 

Furthermore, there is a statistical connection---an anti-correlation---between the width and the edge energy we have not fully explored in our analysis. Given that the increased sharpness of the feature yields a higher energy, this combined with the coarse spectral resolution of \nustar\ (relative to the energy separation of ionized Fe K continua) means the feature could comprise multiple ionized edges (e.g., both \ion{Fe}{25} and \ion{Fe}{26}).  The degree to which each ionization level contributes depends on {\it how} the gas is ionized, i.e., via radiative or collisional processes (see \citealt{Kallman2001}). A full treatment of relative ion fractions is far beyond the scope of this work.

We can use our constraints on the depth of this edge component to estimate the column density of the photoionized gas, given by $N_{\rm Fe 25}=\tau_{\rm edge}/\sigma_{\rm Fe 25}$, where $\sigma_{\rm Fe 25}$ is the \ion{Fe}{25} cross-section, which can be derived from tables provided by \cite{Verner1996}. With $\sigma_{\rm Fe 25}=1.956\times10^{-20}~{\rm cm^2}$ and $\tau_{\rm edge}=0.89$ (from Table~\ref{tab:params}), the ionized Fe column density is $N_{\rm Fe 25}=4.6\times10^{19}~{\rm cm^{-2}}$. If we assume Solar abundances, ($A_{\rm Fe}=3.16\times10^{-5}$; \citealt{Wilms2000}), the photoionized gas has a hydrogen column density of $N_{\rm H}=1.4\times10^{24}~{\rm cm^{-2}}$. This means we are seeing reprocessed emission from either one or more (see Section~\ref{subsec:reflection}) photoionized region(s) that is mildly Compton thick but highly ionized, as opposed to a cold obscuring gas. \\


\subsection{Conclusions}

Through our detailed analysis of five \nustar\ spectral observations of the 2021 outburst of \v4641, we find several ionized Fe K emission lines and a smeared ionized Fe K edge. The line fluxes correlate strongly with the modeled intrinsic disk flux, and the smeared edge is stable in optical depth. Several of these features are well described by a single ionized reflection model, but additional ionized emission components remain, i.e., a single-zone, fixed ionization reflection model cannot explain all the features. Therefore we are likely viewing either multiple photoionized regions or a region with spatial variations in ionization. 

\cite{Shaw2022} argued for significant partial obscuration of the central engine of \v4641\ based on both the unfeasibly small inner disk radius derived from the disk continuum fit to their 2020 \nustar\ data, and the requirement for a partial covering to fit the \chandra\ spectrum down to low energies (below $1$~keV). Similar arguments have been made based on modeling of previous observations of \v4641\ \citep{Revnivtsev2002,Maitra2006,Koljonen2020}. We see tentative evidence for obscuration of the disk based on its low intrinsic luminosity (and thus small inner radius, see Figure~\ref{fig:rin}).~On the other hand, while we cannot rule out the presence of a local partial covering, the strong correlation between the disk and line emission fluxes suggests that a significant portion of the disk emission that is reprocessed is also directly visible to us. We have shown, however, that reprocessing occurs within a highly ionized gas, which agrees with previous conclusions that scattering is taking place in either a large scale height disk or a wind \citep{Koljonen2020}, possibly with a spherical geometry \citep{Shaw2022}. 



Further characterization of the details of the continuum and ionized line and edge features in \v4641\ will require future observations with instruments with higher energy resolution alongside comparable broadband sensitivity, e.g., {\it XRISM} \citep{Tashiro2018}. The combination of {\it XRISM}'s Resolve and Xtend instruments would be ideal for a full characterization of the reprocessed emission we have modeled in this work, given the sensitivity out to $\sim25$~keV and the incredible energy resolution ($\sim4.5$~eV at $6$~keV). This combination will allow us to disentangle the complex ionized Fe emission lines and their relationship to the continuum, as well as identify velocity shifts. Measuring such shifts will facilitate further investigation of the smeared edge feature and bring us closer to understanding its origin and connection to the line features. 

\begin{acknowledgements} 
The authors extend their thanks to the anonymous referee for their constructive comments which have helped turn this into a more complete piece of work. 

RMTC and JN acknowledge support from NASA grant 80NSSC22K0330. 

This research has made use of data, software and/or web tools obtained from the High Energy Astrophysics Science Archive Research Center (HEASARC), a service of the Astrophysics Science Division at NASA/GSFC and of the Smithsonian Astrophysical Observatory's High Energy Astrophysics Division. 

This research has made use of ISIS functions (ISISscripts) provided by 
ECAP/Remeis observatory and MIT (http://www.sternwarte.uni-erlangen.de/isis/). 

\vspace{5mm}
\facilities{\nustar\ \citep{Harrison:2013}, HEASARC, HEASoft}

\software{{\tt XSPEC v.12.13.1} \citep{Arnaud1996}, {\tt ISIS v1.6.2-51} \citep{Houck2000}, {\tt XILLVER} \citep{Garcia2010,Garcia2013}, {\tt RELXILL} (v2.3; \citealt{Garcia2014,Dauser2014}).}
\end{acknowledgements}

\bibliographystyle{aasjournal}
\bibliography{references}
%
%
%
%

\end{document}